# Trends in Processor Architecture

Antonio González
Universitat Politècnica de Catalunya, Barcelona, Spain

## 1. Past Trends

Processors have undergone a tremendous evolution throughout their history. A key milestone in this evolution was the introduction of the *microprocessor*, term that refers to a processor that is implemented in a single chip. The first microprocessor was introduced by Intel under the name of Intel 4004 in 1971. It contained about 2,300 transistors, was clocked at 740 KHz and delivered 92,000 instructions per second while dissipating around 0.5 watts.

Since then, practically every year we have witnessed the launch of a new microprocessor, delivering significant performance improvements over previous ones. Some studies have estimated this growth to be exponential, in the order of about 50% per year, which results in a cumulative growth of over three orders of magnitude in a time span of two decades [12]. These improvements have been fueled by advances in the manufacturing process and innovations in processor architecture. According to several studies [4][6], both aspects contributed in a similar amount to the global gains.

The manufacturing process technology has tried to follow the scaling recipe laid down by Robert N. Dennard in the early 1970s [7]. The basics of this technology scaling consists of reducing transistor dimensions by a factor of 30% every generation (typically 2 years) while keeping electric fields constant. The 30% scaling in the dimensions results in doubling the transistor density (doubling transistor density every two years was predicted in 1975 by Gordon Moore and is normally referred to as Moore's Law [21][22]). To keep the electric field constant, supply voltage should also be reduced by 30%. All together would result in a 30% reduction in delay and no variation in power density. If total area of the chip is kept constant, the net result is twice the number of transistors that are 43% faster, and the same total power dissipation.

More transistors can be used to increase the processor throughput. Theoretically, doubling the number of transistors in a chip provides it with the capability of performing twice the number of functions in the same time, and increasing its storage by a factor of two. In practice, however, performance gains are significantly lower. Fred Pollack made the observation long time ago that processor performance was approximately proportional to the square root of its area, which is normally referred to as Pollack's rule of thumb [24]. This is mainly due to the following two reasons. First, the internal microarchitecture of processors: the performance of many of its key components such as the issue logic and cache memories do not scale linearly with area. Second, even if transistors are smaller by a factor of two, this has not resulted in twice the number of transistors per unit of area due to the increasing impact of wires. Having more functional blocks often requires an increase in the number of wires that is super-linear, which increases the percentage of area that must be devoted to them. An example of this is the bypass logic. Having a full bypass among all functional units requires a number of wires that grows quadratically with the number of units.

On the other hand, the 30% reduction in transistor delay has the potential to make circuits 43% faster. However, benefits in practice are lower, due to the fact that wire delays do not scale at the same pace [13], and because of that, they have become the main bottleneck for microprocessor performance. Besides, as processor structures become more complex, a larger percentage of the area must be devoted to wires, as outlined above, so the impact of wire delays become even more severe. As a result, the time spent in moving data around is the main component of the activity performed by current microprocessors.

Putting it all together, the increased transistor density allows architects to include more compute and storage units and/or more complex units in the microprocessors, which used to provide an increase in performance of about 40% per process generation (i.e. Pollack's rule of thumb). The 30% reduction in delay can provide an additional improvement, as high as 40% but normally is lower due to the impact of wire delays. Finally, additional performance improvements come from microarchitecture innovation. These innovations include deeper pipelines, more effective cache memory organizations, more accurate branch predictors, new instruction set architecture (ISA) features, out-of-order execution, larger instruction windows and multicore architectures just to name some of the most relevant. This resulted in a total performance improvement rate of 52% per year during the period 1986-2003 [12] as measured by SPEC benchmarks. However, this growth rate has dropped to about 22% since 2003 [12].

Multiple reasons have contributed to the slowdown in performance improvement. First, supply voltage has not scaled as dictated by Dennard's recipe; in fact, its decrease has been much lower, in the order of 8% per year (15% every process generation, assuming a 2-year process technology cadence), as shown in Figure 1.

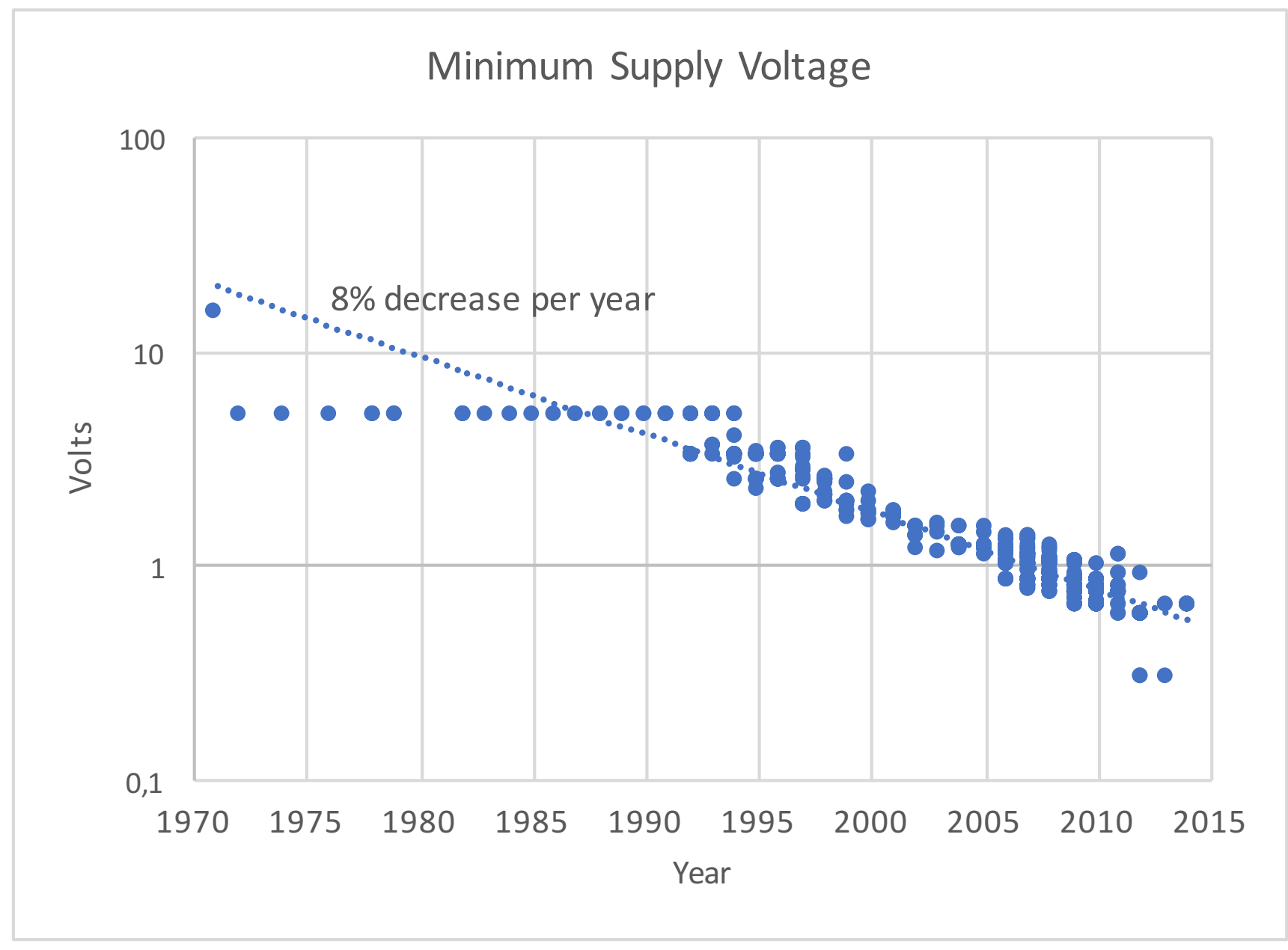

Figure 1: Minimum supply voltage of microprocessors.

The main direct consequence of scaling supply voltage by less than dictated by Dennard's guidelines is an increase in power density and total power of microprocessors. As can be seen in Figure 2, during 1980s and 1990s microprocessor power increased in an exponential manner, by about two orders of magnitude in two decades. An obvious consequence of power increase is an increase in energy consumption and thus in the operating cost of computing systems. More important, this also implied a similar increase in power density since microprocessor area has not changed much over the years. As pointed out by some authors [3], microprocessors in 0.6-micron technology (in the early 1990s), surpassed the power density of a kitchen hot plate's heating coil, and the trend continued increasing by about ten more years, reaching unsustainable levels. Increased power density requires a more powerful cooling solution since power density is a proxy of heat dissipation. Temperature has an important impact on reliability and leakage currents so silicon operating temperature must be kept below a certain limit (in the order of 100 degrees centigrade), which may be unaffordable in some systems due to either its cost or physical characteristics (volume, weight, noise, etc.).

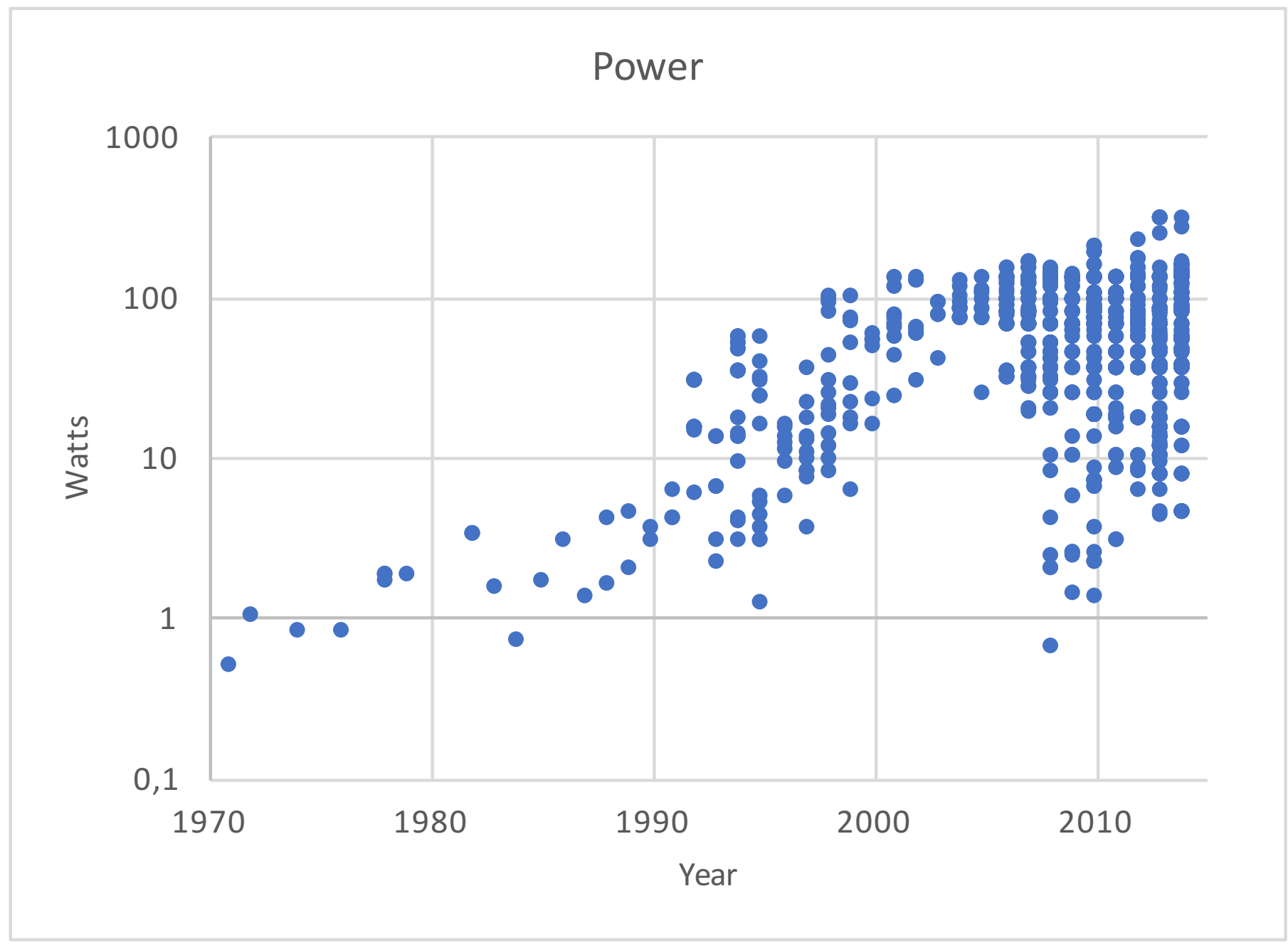

Figure 2: Power (TDP) of microprocessors.

Microprocessor's power growth resulted unsustainable in the early 2000s, and the trend changed towards flat or decreasing power budgets. This explains in part the inflexion point in the performance growth curve which coincides in time with this change in power budget. To keep power density constant when supply voltage scales less than transistors dimensions, the only way is to have a lower percentage of transistors switching at any given time. This motivated the aggressive use of power-aware techniques such as clock gating [34] and power gating [15], which are extensively used by today's microprocessors. Despite the great benefits of these techniques, they have not been able to provide the exponential benefits that would be needed to compensate for the difference between desired and actual supply voltage scaling (30% vs 15%). This has led architects to include techniques to avoid all blocks to be active at the same time, which sometimes is referred to as "dark silicon" [8], or to avoid all blocks to operate at maximum supply voltage through extensive use of dynamic voltage scaling techniques. For instance, in some contemporary processors with multiple cores, when all cores are active supply voltage cannot reach the same level as when only one of the cores is active.

On the other hand, the second important reason that explains the performance inflexion point observed in the early 2000s is due to a slowdown in improvements coming from architecture. During the 1980s and 1990s, there were many innovations in processor architecture, including deep pipelining, out of order execution, branch prediction, data prefetching, superscalar, and memory hierarchy improvements, among others. All these techniques provided significant improvements in performance by either reducing the latency of main functions (especially memory operations) or increasing the degree of instruction-level parallelism (ILP). After two decades of continuous improvements in these areas, they reached a point of diminishing returns. This prompted the inclusion of techniques to exploit thread-level parallelism (TLP) at the processor level, and gave birth to multicore and multithreaded processors, which are extensively used nowadays. TLP techniques have proved to be very effective to increase processor throughput when the workload consists of independent applications. However, they are often less effective when it comes to decompose a single application into a number of parallel threads. Serial parts of the application limit the benefits of TLP, as stated in 1967 by Gene Amdahl [1]. Besides, other important aspects such as workload balance, synchronization and communication cost also pose significant hurdles to the benefits of TLP.

## 2. Current Microprocessors

Figure 3 shows a high-level block diagram of a typical contemporary microprocessor. The main components are a number of general purpose cores (4 in the figure), a graphics processing unit, a shared last level cache, a memory and I/O interface, and an on-chip fabric to interconnect all these components. Below, we briefly describe the architecture of these modules.

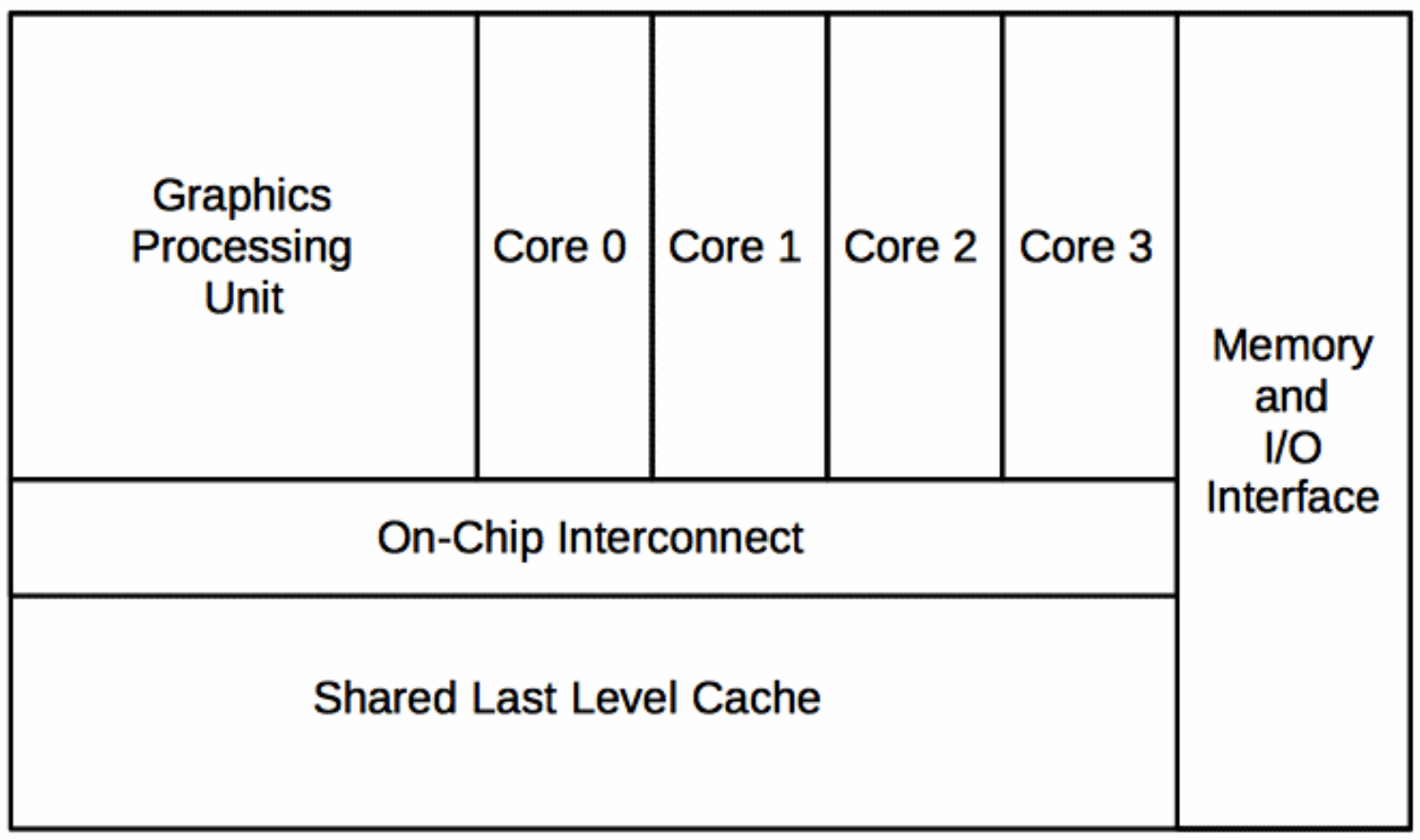

Figure 3: High-level block diagram of a contemporary microprocessor.

## 2.1. General Purpose Cores

The architecture of general purpose cores has significantly evolved throughout its history. From the very early processors, *pipelining* and *cache memories* have been two key microarchitectural techniques used to improve performance.

ILLIAC II [5] and IBM Stretch [2] computers were pioneers in the use of pipelining. Pipelining is an extremely cost-effective technique to exploit instruction-level parallelism, and has been used by practically all microprocessors. A pipeline with N stages can potentially overlap the execution of N instructions and thus, provide an increase in throughput by a factor of N. In practice, benefits are lower due mainly to instruction dependences, which introduce bubbles in the pipeline. Initial processors had just a few pipeline stages but the introduction of RISC architectures in the early 1980s made pipelining to be more cost-effective and facilitated the use of deeper pipelines. Innovations in microarchitecture techniques to avoid pipeline bubbles due to dependences, such as out of order execution, data dependence speculation and branch prediction, allowed a continuous increase in pipeline depth, during the 1980s and 1990s. Some studies in the early 2000s suggested that the optimal pipeline depth for performance could be in the order of 50 stages [14][28], but some later studies [10][29] showed that when power is taken into account, the optimal pipeline depth is much shallower, and in practice, pipeline depth increase stopped at around 20 stages. Energy-efficiency favors shallow pipelines since deeper pipelines consume more energy due to extra latching, more complex control and energy waste due to speculative execution techniques. As a result, current microprocessors pipelines normally have between 10 and 20 stages and the forecast for the future is that pipeline depths will keep around this range.

Microprocessors require many accesses to memory to execute a program. Since instructions and data are stored in memory, a typical instruction requires four memory accesses, one to fetch the instruction, two to read its two source operands, and one to write its result. While processor performance has increased at a rate ranging from 22% to 52% per year, as commented above, memory latency has improved at a much lower rate of about 7% per year. In a period of 40 years, this resulted in a gap of four orders of magnitude [12]. This trend motivated the organization of memory in a hierarchical way. A hierarchical organization exploits a core feature of the technology: smaller blocks are normally faster. This applies to all kind of logic blocks, and in particular to memories: a small memory is faster than a larger one. This is basically due to the fact that latency of read and write operations are dominated by wire delays of the word and bit lines. The smaller the memory, the shorter word lines and bit lines are.

A hierarchical organization of memory was used as early as the mainframes of the 1960s, which used a hierarchy of physical memory composed of modules built with different technologies: semi-conductor, magnetic core, drum and disk. Data caches were used for the first time in the IBM 360 Model 85 computers [16]. Most computers nowadays have a memory hierarchy organized in six levels: a register file, three levels of cache memory, main memory and disk.

Another important microarchitecture technique used since the very early processors has been *branch prediction*. When a branch is fetched, the next instructions to be executed depend on the branch outcome, which is not available until the branch is executed some cycles later. Stalling the fetch until the outcome of the branch is known would cause a significant penalty in pipelined processors since branches are very common in many programs (in the order of one out of ten instructions for non-numerical codes). Predicting the outcome of the branch and speculatively executing the instructions of the predicted path can alleviate this penalty provided that the predictor is highly accurate. This technique is known as branch prediction and was used as early as the late 1950s in the IBM Stretch computer [2].

Branch predictors based on a table of two-bit saturating counters were introduced by James Smith in the late 1970s [26] and have been used in practically all microprocessors since then. Since the penalty of a branch misprediction grows linearly with the pipeline depth, deeper pipelines prompted research on more sophisticated predictors such as the two-level branch predictor proposed by Yeh and Patt in the early 1990s [35], variants of which have been used by many modern microprocessors. These predictors are based on two-bit saturating counters, but the particular counter used in each case depends not only on the particular branch being predicted but also on the outcome of recent past branches. Another important innovation in the area of branch prediction that has been used by many microprocessors is the concept of hybrid predictors, first introduced by Scott McFarling in 1993 [20]. The key idea is based on two observations: first, the most cost-effective branch predictor is different for different types of codes; second, more sophisticated predictors tend to have a longer warm-up time, so at the beginning of each context-switch interval simple predictors tend to be more accurate than complex ones, whereas this trend is inverted once the complex predictor has been warmed-up. Based on these observations, he proposed a mechanism to combine multiple predictors and dynamically choose which one is more likely to be the most accurate.

The instruction set architecture (ISA) is another important feature of a processor architecture. The ISA has evolved over the years to include new instructions that can better support common code constructs in mainstream applications domains. An important change in the design of ISA was introduced in the early 1980s, mainly driven by independent projects at UC Berkeley [23], IBM [25] and Stanford University [11]. This trend was coined as *RISC* (Reduced Instruction Set Computer) by the team in UC Berkeley, term that was later adopted by everyone else. Up until then, the trend in ISA design was to increase its complexity. It was believed that by closing the semantic gap between high-level languages and machine languages, computer could be made to be more efficient. These projects demonstrated that a much simpler ISA could be competitive, or even outperform, a more complex one, and at the same time offer additional advantages such as a much lower cost of design and verification of the processor. The maturity of compiler technology combined with the benefits of circuit simplicity in terms of delay, cost and energy consumption were some of the key aspects that favored the success of RISC ISAs.

Enhancements in pipelining, amplified by the simplicity introduced by RISC ISAs, caching and branch prediction were the main driving forces behind microarchitecture innovation during the 1980s. These processors aimed to achieve a throughput of one instruction per cycle, which they approached after several generations of improvements. During the 1990s, a new microarchitecture organization to exploit further ILP became common use, which was coined *superscalar processors*. A superscalar processor is a processor that can process multiple instructions in all its pipeline stages. That is, it can fetch, decode, rename, issue, execute and commit multiple instructions at the same time. How many instructions can be processed in parallel in each pipeline stage is referred to as the width of the superscalar processor. In practice, not all pipeline stages may have the width, so we define the width of the processor as the minimum width of its pipeline stages. A N-wide superscalar processor can potentially achieve a performance of N instructions per cycle.

A superscalar processor requires to replicate hardware resources, normally by the same factor as its width, but in some parts of the processor, the hardware cost may grow superlinearly with the width. An example of this superlinear cost is the bypass logic, which grows quadratically with the number of functional units. Another not so obvious example is the issue logic. In this case, the reason is that to find more independent instructions to be issued every cycle, the hardware has to search in a larger instruction window, so if we double the issue width, we need twice the number of issue ports, but each port has to search in a much larger window. Superscalar width grew during the 1990s, but due to this superlinear cost, and the difficulties to find enough ILP, it flattened out to around 4, and practically all current processor have a superscalar width between 2 and 8.

Another key microarchitecture technique that became common use during the 1990s was *out-of-order execution*, also known as dynamic scheduling. The main idea is to allow the hardware to execute the instructions in an order different to the one that they appear in the binary, with the constraint that the semantics of the program must not be changed. The goal is to find more ILP by reordering the instructions. For instance, the consumer of a load that misses in cache may be stalled during many cycles. Instead of stalling all instructions younger than this consumer, as in-order processors do, out-of-order processors can execute younger instructions, provided that they do not depend on the load. Out of order execution provides important benefits in performance but has also an important cost, mainly due to a much more complex issue logic, including the issue logic of memory instructions, which normally is separated from the issue logic of the rest of instructions, and is quite costly since memory dependences are more difficult to check than register dependences. This is due to the fact that register dependences are known at decode time and can easily be identified as instructions are decoded/renamed in program order, whereas memory dependences need to be identified later in the back-end of the pipeline, once the effective address of loads and stores are computed, and this computation is performed out of program order.

Out-of-order execution became popular in the 1990s and is used by the vast majority of current microprocessors. However, the main ideas behind this technique date back to the 1960s. In fact, out-of-order execution was first introduced by the CDC 6600 computer [31] in 1964 and was later improved by the IBM System/360 Model 91 [32]. IBM approach is usually referred to as Tomasulo's algorithm, and is the bases for the schemes used nowadays. The main innovation introduced by IBM over the CDC scheme was the use of *register renaming*. By renaming register operands, the processor has more options to reorder the code, since it only needs to respect data dependences (a.k.a. read-after-write dependences). Name dependences (a.k.a. write-after-write and write-after-read dependences) are removed and do not impose any ordering constraint. Register renaming provides huge benefits in terms of ILP and thus, is used by all current out-of-order processors to the best of our knowledge.

During the 2000s, the main architectural innovation introduced at the microprocessor core level is called *multithreading*. There are different variants of the concept of multithreading, but the most commonly adopted by current microprocessors is known as *simultaneous multithreading* [33] and it was first used by commercial processors in the Intel's Pentium 4, under the name of hyperthreading [17]. The key idea behind this technique is to provide a microprocessor core with the capabilities to execute multiple threads simultaneously sharing the majority of hardware resources. At any given cycle and any given pipeline stage, the processor can potentially execute multiple instructions belonging to different threads. In this way, the processor can remove some pipeline bubbles and the multiple slots of a superscalar processor in each pipeline stage can be more frequently filled with useful work. In other words, multithreading increases the utilization of the resources already present in a pipelined, superscalar processor.

This technique is highly efficient from the hardware cost point of view, since it requires small extensions to a conventional superscalar processor. Its main cost is an increase in register file storage, since the processor has to keep the state of multiple threads simultaneously. On the other hand, for this same reason, its scalability is limited. Since practically all processor core resources are shared among the simultaneously running threads, there are frequent structural hazards that prevent each thread to run at the same speed as it would if it run alone. The benefits of increasing the degree of multithreading diminish quickly for general purpose cores. With two to four threads, resources get highly utilized, and adding more threads would bring minimal benefits in many cases. Because of that, the typical multithreading degree of general purpose cores is in the range of two to four, although for server processor it may be a bit higher, such as the recently announced IBM Power9 [30] that supports up to 8 simultaneous threads. Server workloads are sometimes highly memory intensive, and the utilization of the CPU resources is low, which allows the possibility to exploit a larger number of simultaneous threads. Another type of processing unit that is highly multithreaded is the graphics processing unit (GPU). In this case, GPUs are designed to support a very large number of threads since graphics applications exhibit a huge degree of TLP (e.g., most of the operations needed to compute the color of each pixel are performed by independent threads). GPUs rely on a high degree of multithreading to hide memory latencies rather than on a complex memory hierarchy. Even if threads are frequently stalled due to long latency memory accesses, there are normally other threads ready that can use the hardware resources to make progress and keep the hardware highly utilized.

To conclude this section, Figure 4 shows a high-level overview of the typical microarchitecture of modern general-purpose cores.

Instructions are fetched from an on-chip instruction cache, which normally has a few tens of kilobytes. The main component of the instruction fetch logic is a branch predictor. Multiple instructions can be fetched in the same cycle by using a single, wide cache memory port that can provide multiple consecutive bytes in a single access. Branch prediction is performed in parallel with the instruction cache access (or even before in some processors) to avoid bubbles in the pipeline (otherwise, in the next cycle the processor would not know which are the next set of instructions to fetch). Multiple fetched instructions can potentially be branches, so the branch predictor needs to predict multiple branches in parallel. However, depending on the branch predictor being used this may not be that easy. In particular, if the branch predictor uses global history, when predicting a branch the outcome of the previous branches must be known. This is typically solved by assuming that the previous branches being predicted in parallel are not taken. In case any of the branch predictions turns out to be taken, all the following branch predictions are discarded. Since all instructions fetched in parallel are consecutive, all instructions after the predicted taken branch are discarded too, since they are not in the predicted path. Thus, the branch predictor can normally predict multiple branches per cycle but only up to one taken branch.

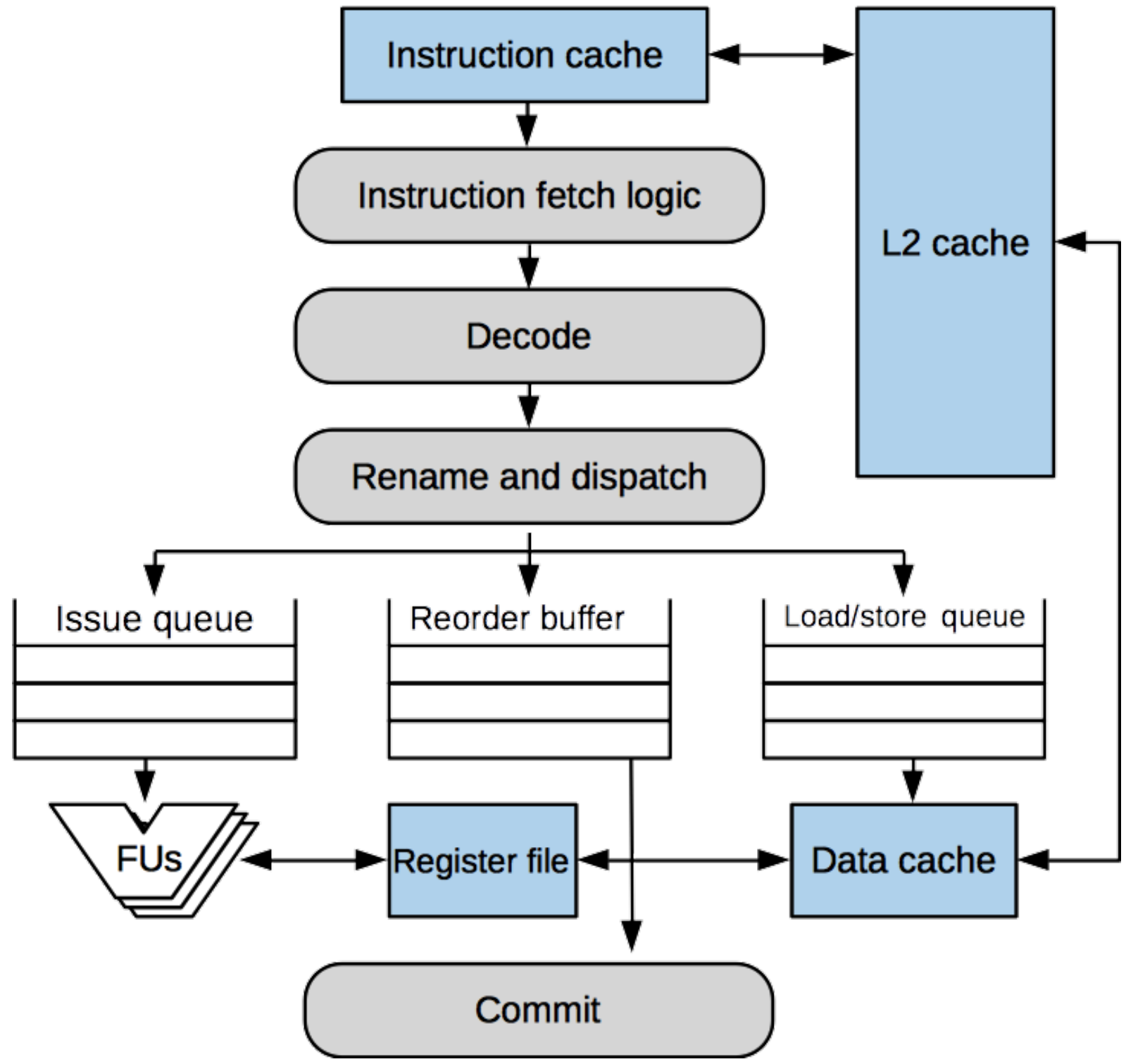

Figure 4: Microarchitecture of a contemporary general-purpose core.

After being fetched, instructions are processed by the decode logic, which identifies their type, which operands are required and other relevant fields of the instruction. Decoding multiple instructions in parallel does not pose any challenge and simply requires to replicate hardware since each instruction can be decoded independently of the others. Some processors perform a dynamic translation at this stage. This is common in processors using old ISAs such as x86, in order to translate the compiler-generated instructions to an internal ISA which is more amenable for pipelining and other microarchitectural optimizations. This translation is relatively simple since it is done for each instruction individually; optimizing blocks of instructions would be more effective but also much more complex and is normally not used.

Afterwards, instructions rename their register operands. As outlined above, this step is used by out-of-order processors to remove name dependences and increase in this way the amount of ILP. In-order processors normally do not make use of it. Renaming is done basically through a table that keeps track of the latest storage location (a.k.a. physical register) assigned to each logical register. Source operands are translated to physical registers by reading the corresponding entries in the renaming table. The destination register of each given instruction is assigned to a free physical register (if there are no physical registers available, renaming is stalled). Superscalar processors rename multiple instructions per cycle. This requires a rename table with multiple read/write ports, and some additional logic to take into account data dependences among instructions being renamed in parallel. In particular, each source operand must be checked with all previous destination operands, and in case of a match, the physical register identifier used by this source operand is not the one coming from the renaming table but the one assigned to the closest matching destination operand. Next, instructions are dispatched to the issue queue and the reorder buffer, and memory instructions are also dispatched to the load/store queue.

All the above operations are performed in program order in all processors, including out-of-order processors. This part of the pipeline is normally referred to as the front-end. An in-order front-end simplifies the detection of register data dependences among instructions, but on the other hand, exacerbates the penalty of any potential stall. Most harmful stalls in the front-end pipeline are due to instruction cache misses.

Instructions remain in the issue queue until they are selected to be executed. The process to select the instructions to be executed every cycle is called instruction issue. For in-order processors this logic is relatively simple. To issue N instructions per cycle, it just needs to check the N oldest instructions in the issue queue. Each of these instructions is issued if its source operands are ready, the required resources for its execution are available (e.g., functional units, register file ports, etc.) and all older instructions are issued too.

Most processors nowadays can issue instructions out-of-order. This is much more complex since it requires to check all instructions in the issue queue, since all of them are candidates to be issued. To this end, the issue queue stores the renamed source operands of each instruction and a ready bit for each of them. Every time an instruction is executed, its destination register id is checked against all source operands in all entries, and for every match, the corresponding ready bit is set (this action is normally referred to as wake-up). Instructions that have all source operands ready, and the required resources are available are selected to be executed at every cycle. If there are more candidate instructions than available resources, a heuristic is applied to prioritize the candidate instructions. Given priority to the oldest instructions is quite common, and sometimes this is combined with other simple heuristics such as prioritizing long latency instructions.

Each issued instruction reads first its source operands, then is executed in a functional unit and finally writes its result in the register file. The register file is therefore highly multi-ported, to support multiple reads and writes per cycle. For instance, a 4-way issue processor may typically have 8 read ports and 4 write ports. Processors have multiple functional units to execute multiple operations in parallel, including integer ALUs and floating-point ALUs. Besides, practically all processors include a rich set of SIMD (single-instruction, multiple-data) instructions in their ISA. These are instructions that operate on vector operands stored in registers, and are usually known as multimedia extensions, since they are very effective for multimedia applications. To support these instructions, processors also include SIMD units, which can process all the elements of a vector operation in parallel. Since there are three different types of register operands, i.e., integer, floating-point, and vector, some processors have three separate register files, one for each type, whereas others have only two, one for integer and another for floating-point and vector.

Instructions wait in the reorder buffer until they commit. The reorder buffer stores some bookkeeping data for each in-flight instruction plus the information required in case the instruction needs to be squashed. An in-flight instruction may be squashed for a variety of reasons such as an older branch being mispredicted, or an older instruction generating an exception. Instructions commit in program order, even if the processor is out-of-order, since an easy way to guarantee that an instruction will not need to be squashed is waiting for the commit of all older instructions. At this point, no older instruction may cause an exception, branch misprediction nor any other type of event.

Memory instructions deserve special attention. To be issued, they must go through a similar process as other instructions. However, the actions to be performed are a bit more complex since dependences are more difficult to identify because they involve memory locations rather than registers. Register operands' ids, and therefore register dependences, are known during renaming, but memory addresses, and thus memory dependences, are not. The address of the memory location read or written by a memory instruction is normally computed at run time by adding an offset to the content of a register. Until this register operand is not ready, the address cannot be known. The logic associated with the load/store queue is responsible for computing these addresses, checking for potential dependences and deciding when memory reads can be issued.

Store instructions do not write into memory until they commit. This is because a write to memory cannot be undone, unlike writes to registers, since they may be visible to other processes. Once a write is visible to another process, it may trigger some activity in this process (e.g., entering a critical region), which could not be undone if the store turned out to be squashed.

A load instruction can safely read from memory as soon as its effective address has been computed and there are no dependences with older stores. Checking for dependences requires to compare memory addresses, which are much longer than register ids, and besides the check has to take into account the potential different data widths of loads and stores.

To guarantee that a load does not depend on any older store, it must wait until the effective address of all older stores are computed. This may significantly delay a load, and most of the times this delay is unnecessary since memory dependences with instructions close enough to be in the instruction window at the same time are not common. To avoid this delay, some processors use some kind of mechanism to predict memory dependences and just stall loads that are likely to have a dependence. This family of mechanisms is called data dependence speculation [9][19], and it requires a recovery scheme for mispredicted loads, similar to the one used for branch mispredictions.

Store-to-load forwarding is commonly used when a memory dependence is encountered/predicted between a load and a not committed store. Rather than waiting for the store to write to memory, the load gets the data directly from the store.

## 2.2. Multicore Processors

The vast majority of current microprocessors have multiple general purpose cores based on the architecture described in the previous section. Figure 5 depicts the main components of a multicore processor. There are a number of cores, each one with private L1 caches (separate caches for instructions and data) and a private second level cache (some processors do not have a private L2 cache), a shared last level cache and an interconnection network that allows all the cores to communicate through the memory hierarchy. The lower levels of the memory hierarchy are normally a main memory and a disk storage, and they are located off-chip. A multicore processor can run multiple threads simultaneously in different cores with no resource contention among them except for the shared resources, which are basically the memory hierarchy and the interconnection network. The architecture of these two components is key for the performance of multicore processors and are described in more detail below. The architecture of each one of the individual cores is basically the same as in a single-core processor, and has been described in the previous section.

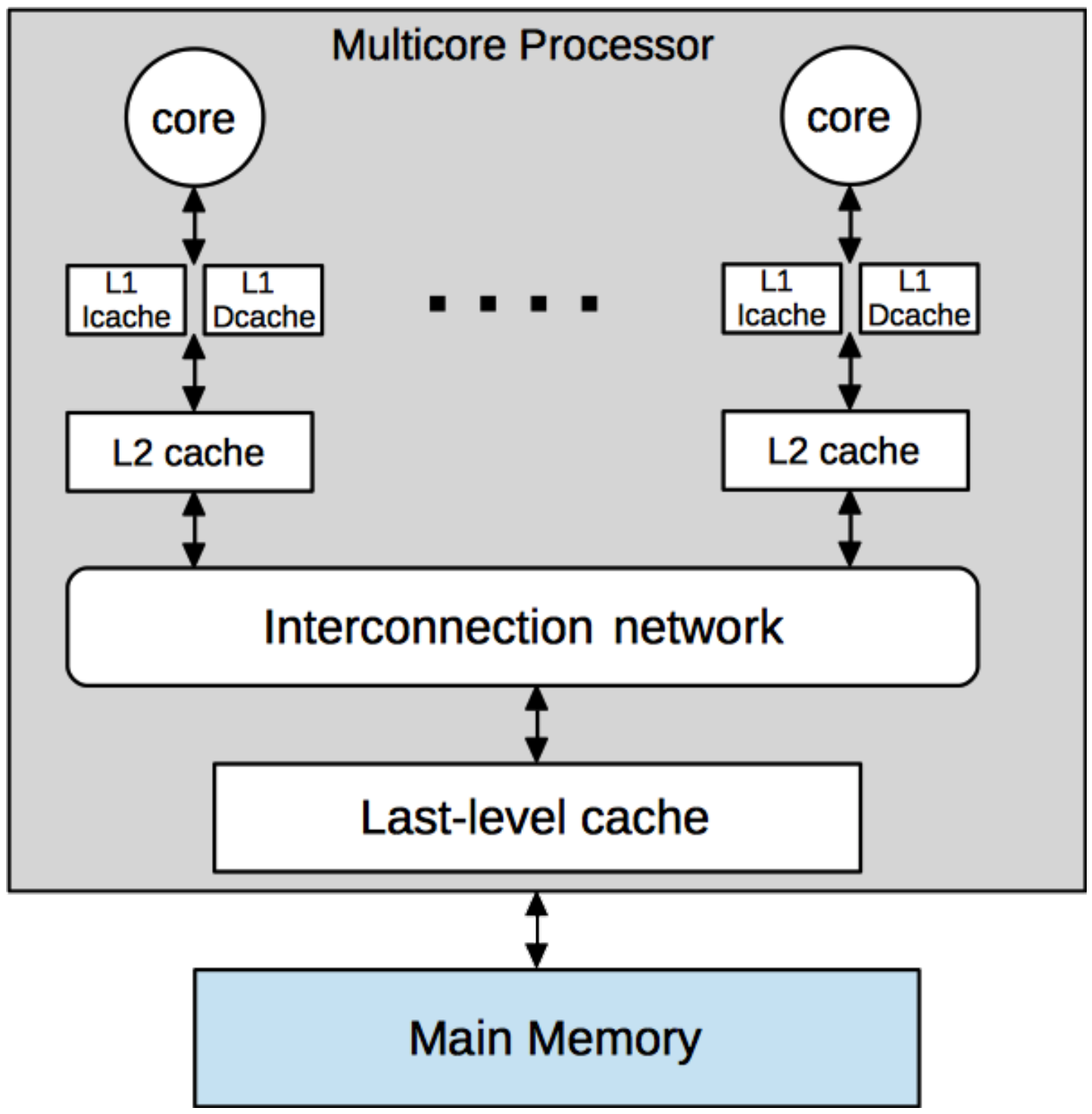

Figure 5: Block diagram of a multicore processor.

### 2.2.1. Memory Hierarchy

Parallel threads running in different cores can share the memory address space, which is a very desirable feature for programmability. However, this introduces two important challenges: how to keep data coherent when it is used by multiple threads in different cores, and how to make memory operations visible to the other threads running simultaneously in other cores. The former challenge is solved by a new mechanism that is called *cache coherence*. The latter is handled by what is called the *memory consistency* model. We briefly describe these two components below.

Private caches are great to improve the performance of each individual core, but they introduce some problems when accessing variables that are shared among multiple cores. The problem arises when multiple threads running in different cores access the same variable and at least one of them performs a write operation. Since variables are normally copied in the private caches, a write performed by a given core would not be visible to the other cores, since they would keep seeing the value store in their respective private caches. If the cores did not have private caches and all read and write operations were performed in a completely shared memory hierarchy, this problem would not exist. Cache coherence is a microarchitecture mechanism that solves this problem by making private caches transparent to the programmers. In other words, with cache coherence, a processor with private caches produces exactly the same results as the same processor with all private caches removed.

Cache coherence is implemented through a coherence protocol. The basic idea of a coherence protocol is to make sure that writes from a given core are propagated to all potential copies of the same variable in the memory hierarchy, including private caches of other cores. There are two main families of protocols which are called write-invalidate and write-update protocols respectively, the former being the more common solution. In a write-invalidate protocol, when a core has to perform a write operation, all copies of the same data elsewhere are invalidated before performing the write. If other cores are using this variable and have a private copy, they will have to request it again after being invalidated, and will get the updated value from the core that performed the write. In write-update protocols, a write operation updates all copies of the variable in the system. They are more complex than write-invalidate protocols, and consume more network bandwidth thus, they are rarely used.

The implementation of a cache coherence protocol is complex, and requires a careful design and a thorough validation to guarantee that it is correct. The complexity comes from the fact that a transaction (e.g. an invalidation request) generated by one core has to be propagated to the other cores and this cannot not be done instantaneously. In the meantime, while a transaction is being processed, other request to the same variable may occur in the system, and the system has to deal with them and guarantee correct semantics and no deadlocks. This is normally solved by having a number of stable states that represent the state of a memory block after a transaction has been completed plus a number of transient states that represent the state in different phases of an on-going transaction. The memory block granularity normally used to keep state information is a cache memory block. The most typical stable states used are the following:

- Invalid: the block is not present or is stale.
- Shared: The block can be read but not modified.
- Exclusive: The block can be read, and it is the only valid copy.
- Modified: The block can be read and written.

There are two main families of protocols regarding how the state information is stored and how transactions are processed. In a snoopy-based protocol, a transaction generated by a cache controller is broadcast to all other controllers. The system relies on the network to guarantee that all messages arrive to all cores in a consistent order. For instance, most protocols assume that they arrive in the same order to all cores, and use a shared bus to guarantee this order. The other family of protocols is called directory-based. In this case, there is a directory that holds the state of each memory block and normally sits next to the LLC. A transaction generated by a cache controller is sent to the directory, and based on the information in the directory, the request is served by the LLC or it is forwarded to the corresponding private cache(s). In general, snoopy-based protocols are simpler but directory-based protocols are more scalable and thus, more effective for a large number of cores.

The memory consistency model is a formal definition of which potential executions (i.e., outcomes) of a parallel program are allowed and which ones are not. Note that multiple different outcomes for a given program may be allowed since many parallel programs are non-deterministic. In other words, the memory consistency model specifies what values load instructions may return and what is the final state of the memory for a given input data. This information is needed by programmers in order to write programs that do what they expect, so it is part of the ISA.

For instance, assume a code in which two threads modify a different variable each, and later they read the variable modified by the other thread. If there are no synchronization operations in between, multiple outcomes are possible. The thread that executes ahead of the other will get the old value, whereas the other thread will get the updated one. Alternatively, both threads can end up reading the updated value if the two threads perform the two writes before the two reads. All systems allow these three alternative outcomes but some systems also allow a less intuitive one, which is that the two threads read the old value. The memory consistency model precisely defines which of these four outcomes is allowed, so that the programmer knows what to expect when the code is executed.

Different memory consistency models represent a different trade-off between programmability and performance. The most intuitive consistency model for programmers is called *sequential consistency* and was introduced by Lamport in 1979 [18]. A system is sequentially consistent if "the result of any execution is the same as if the operations of all processors (cores) were executed in some sequential order, and the operations of each individual processor (core) appear in this sequence in the order specified by its program". This memory model is the most intuitive for programmers but is also the one that imposes more constraints to the hardware in terms of when memory accesses can be performed.

Another widely used memory model nowadays is called total store ordering, which is used by SPARC and x86 processors [27]. For most program idioms, it behaves like sequential consistency, but it allows some new executions not permitted by sequential consistency. For instance, for the code example above, total store ordering allows both threads to read the old value, whereas sequential consistency does not. More precisely, total store ordering imposes the same ordering constraints as sequential consistency with one exception: loads can be moved above older non-conflicting (to different addresses) stores. The motivation for removing this constraint is performance. By removing it, the processor can use a write buffer to hold committed stores until they get read-write permissions. This buffer is very effective to hide the memory latency for store misses, so it results in significant performance improvements, and is widely used by microprocessors.

### 2.2.2. Interconnection Network

As illustrated in Figure 5, a multicore processor consists of a number of cores and a last level cache that need to communicate among them. This communication is carried out through an interconnection network. In general, each core and its private caches are a node in this network, and the last-level cache is another node of the network. It is also common that the last-level cache is split into multiple modules, each one being a different node of the network. Each one of these nodes is connected to a switch through a network interface. Each switch has a number of links that connect it to other switches as shown in Figure 6.

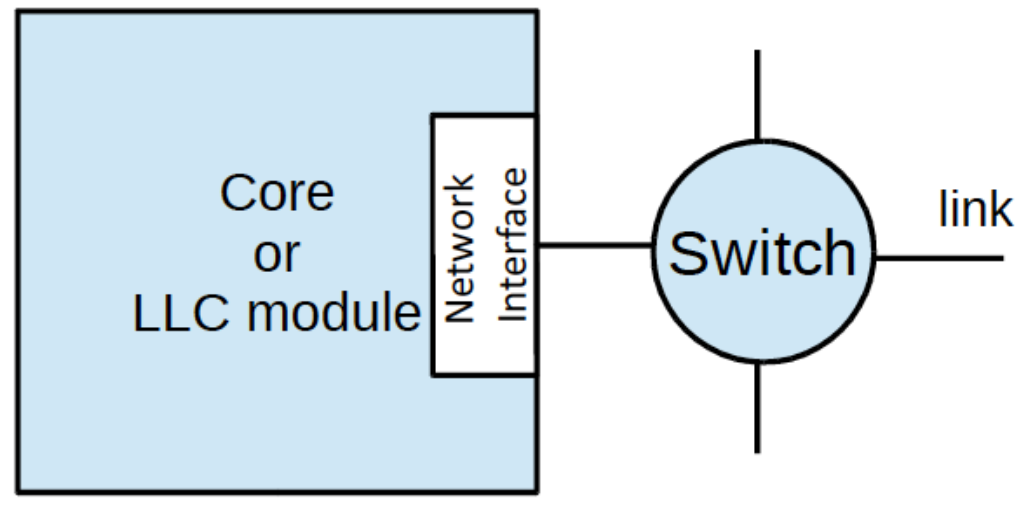

Figure 6: Block diagram of network node.

The different switches are interconnected following a particular topology. For instance, Figure 7 shows a mesh topology, which is one of the most commonly used topologies for multicore processors. A line is another common topology, which is basically a mesh with just one node in one of the two dimensions. If the two nodes at the end of the line are interconnected, then we have a ring for the one-dimensional case or a torus for a two-dimensional case, which are also commonly used.

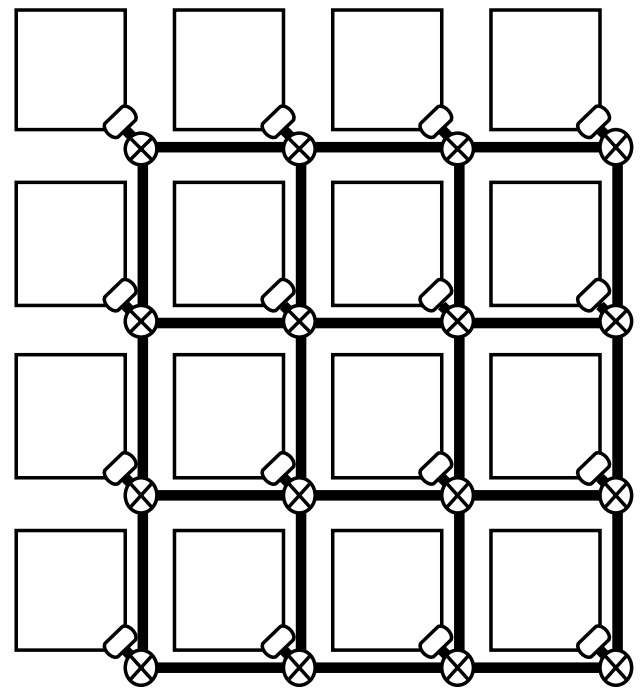

Figure 7: A mesh network topology.

## 3. Going Forward: Specialized Units

We are witnessing a slowdown in process technology improvements. Supply voltage has not scaled as much as Dennard's guidelines [7] for years and Moore's Law scaling [21][22] has slowed down in recent years and is approaching a limit unless some breakthrough in process technology appears in the coming years. Chip manufacturing technology has been based on silicon for decades. Current process technology uses a 10nm fabrication process and the minimum feature size of circuit components is approaching atomic dimensions. Taking into account that silicon lattice spacing is around 0.5 nm, the current 10nm feature size is equivalent to around 20 atoms. Further miniaturization is going to be very difficult to achieve and dimension scaling will very likely stop in a few more generations.

On the other hand, increasing energy-efficiency will keep being a main driving force for innovation in computing systems. This is easily explained by observing that the power dissipation of a system is equal to the average energy consumed per task multiplied by the number of tasks that the system can performed by unit of time. Note that the latter term is the performance of the system whereas the former is what we call energy efficiency. If power cannot be increased due to the various reasons explained in the section 1 of this chapter, any increase in performance requires a reduction in energy per task of the same magnitude. In other words, the challenge for future computing systems is not only how to make them run faster, but at the same time, this increase in performance has to come with a similar decrease in energy consumption (i.e., a similar improvement in energy efficiency). This is really challenging, since normally higher throughput implies higher energy. For instance, a faster addition requires to increase the supply voltage of the adder or to use a more sophisticated adder. In both cases, the energy to perform an addition increases.

Since process technology is unlikely to provide significant additional benefits in the future, the improvements in energy efficiency have to come from other areas such as microarchitecture. However, the microarchitecture of general purpose cores has been optimized for several decades, and there is little headroom for further improvements. Perhaps the most promising approach is "specialization". It is well known that a specialized unit can provide dramatic improvements in energy efficiency when compared to a general-purpose unit. Specialized units are already a reality in most processors. For instance, practically all current processors include a graphics processing unit, which is a unit specialized on image rendering. Most processors for smart phones include specialized units for image and audio processing.

In the future, we are likely to see an explosion in the use of specialized units. The drawback of specialization is its higher cost in comparison with general-purpose units, since their cost needs to be amortized over a smaller number applications. However, the flip side of the death of Moore's Law will be a significant decrease in cost of chip fabrication, since manufacturing industries will not need to replace their equipment so often, and their investments in process technology will be drastically reduced. Under this scenario, transistors will be very chip, much chipper than they already are, and this will open the door for many new opportunities to use them in specialized units.

## Acknowledgements

This work has been supported by the CoCoUnit ERC Advanced Grant of the EU's Horizon 2020 program (grant No 833057), the Spanish State Research Agency (MCIN/AEI) under grant PID2020-113172RB-I00, and the ICREA Academia program.